# Preprint notes

Title of the article:
Ethically Aligned Design of Autonomous Systems: Industry viewpoint and an empirical study
Authors:
Ville Vakkuri, Kai-Kristian Kemell, Joni Kultanen, Mikko Siponen, Pekka Abrahamsson

Notes:
- This is the author's version of the work
- The definite version is submitted to journal track of Transport Research Arena, (TRA2020) for peer review

# Ethically Aligned Design of Autonomous Systems: Industry viewpoint and an empirical study

Ville Vakkuri[a]*, Kai-Kristian Kemell[a], Joni Kultanen[a], Mikko Siponen[a], Pekka Abrahamsson[a]

[a] *University of Jyväskylä, P.O. Box 35, Jyväskylä FI-40014, Finland*

**Abstract**

Progress in the field of artificial intelligence has been accelerating rapidly in the past two decades. Various autonomous systems from purely digital ones to autonomous vehicles are being developed and deployed out on the field. As these systems exert a growing impact on society, ethics in relation to artificial intelligence and autonomous systems have recently seen growing attention among the academia. However, the current literature on the topic has focused almost exclusively on theory and more specifically on conceptualization in the area. To widen the body of knowledge in the area, we conduct an empirical study on the current state of practice in artificial intelligence ethics. We do so by means of a multiple case study of five case companies, the results of which indicate a gap between research and practice in the area. Based on our findings we propose ways to tackle the gap.

* Corresponding author. Tel.: +358-41-515-0101;
   *E-mail address:* ville.vakkuri@jyu.fi

# 1. Introduction

*Artificial Intelligence* (AI) systems are becoming increasingly ubiquitous. Most inhabitants of the developed world already interact with various AI systems on a daily basis, perhaps even without realizing it themselves. The more sophisticated recommendation systems utilized by various B2C Software-as-a-Service media platforms such as YouTube or Netflix utilize AI and *Machine Learning* (ML), and specifically Deep Learning (Lecun et al. 2015), to generate personalized recommendations for their users. *Autonomous Vehicles* (AVs) operated by AI are slowly entering the public roads, and AI-based surveillance systems armed with facial recognition capabilities are already being deployed (Conley et al. 2014). AI systems have also been explored in the field of medicine given e.g. their ability to aid in providing diagnoses by processing data (Hamet and Tremblay 2017). In general, progress in the field of AI has been faster than anticipated by many, experts (MIT Technology Review 2017) and the general public alike.

One key difference between AI, and specifically autonomous systems, and conventional software systems is that the idea of an active user sometimes becomes blurred. One seldom *uses* AI systems as opposed to being an object to their data collection procedures or other actions. While you can simply not use a conventional software service if you do not like it, you have little control over whether a company filters your job application using an AI. Moreover, many AI systems are *Cyber-Physical Systems* (CPS) that operate both in the digital world and the physical world as opposed to being purely digital, conventional software systems. CPSs are numerous, ranging from security cameras to cargo ships, and exhibit various degrees of autonomy. While most AI-based CPSs until recently operated in highly confined settings (e.g. factory robots), CPSs such as AVs are now entering public spaces where they can interact with unassuming passers-by (Charisi et al. 2017). These systems are particularly notable from the point of view of ethics as they have clear physical damage potential (Charisi et al. 2017).

Given their potentially enormous societal impact, AI systems should be designed while taking ethics into consideration (Bostrom and Yudkowsky 2018; Bryson and Winfield 2017). For example, when an AV gets into an accident, we should always be able to understand *why*. This is not always simple even with full access to the program code as ML systems can be highly complex even to their creators (Ananny and Crawford 2018), given the nature of ML where the AI learns by going through vast masses of data. Another factor that makes ethical consideration challenging at times is that the effects of the systems are not always direct. An individual AV that crashes has a clear effect, but systemic effects resulting from swarms of AVs driving on the roads are more difficult to evaluate (e.g. human actors may drive more carefully or even irrationally near AVs out of caution, confusing the systems). Often, however, ethical aspects are seemingly not even considered in the first place as incidents such as the recent Cambridge Analytica case perhaps underline.

Though the academic discussion on AI ethics has recently accelerated, the state of the arts on the field remains largely unknown. In this paper, we look at the current mindset in the industry in relation to Ethically Aligned Design (EAD) of AI systems. Through a case study of six cases, we seek to understand whether ethics in AI are considered important or even relevant at all in the industry. The specific research questions in this paper are thus formulated as follows:

> **RQ:** What practices, tools, or methods, if any, do industry professionals utilize to implement ethics into AI design and development?

The rest of this paper is structured in the following manner. The second section further discusses AI, specifically in relation to autonomous CPSs and transportation. In the third section we discuss our research methodology before presenting our findings in the fourth section. We then discuss the implications of the findings in the fifth section before concluding the paper in the sixth and final section.

# 2. Theoretical Framework

*2.1. Background: Artificial Intelligence and Autonomous Vehicles*

Currently, AVs are being developed across industries. Though arguably the most media exposure is on cars given their nature as B2C personal vehicles, the possibilities of AI have been explored in relation to drones, cargo ships, buses, trains, and airplanes alike. Few examples (e.g. SVT Nyheter 2019) of fully autonomous vehicles in public use exist at present, although the degree of autonomy exhibited by various types of vehicles is steadily increasing.

Regardless of software quality in AVs, accidents and dangerous situations are arguably inevitable (e.g. resulting from a sensor fault (The Guardian 2016)). However, whereas human actors seldom have time to make an informed decision, or sometimes even react, in the face of an impending accident, AI systems are capable of making a decision near instantaneously. While any autonomous CPS should obviously seek to preserve human lives whenever possible, such a system may end up in a dilemma, such as the commonly cited example "Should Your Car Kill You To Save Others?" (Popular Mechanics 2016).

In addition to the notion of material damages, ethics in relation to data handling should always be considered in AI systems (Flores et al. 2016). AI systems work on data and machine or deep learning based systems are trained using huge sets of data. When used in training systems, data can result in bias or simply unwanted learning. For example, Amazon noted that its recruitment AI turned out to be highly biased against women as a result of being trained using data of their past hires which had been predominantly male (Reuters 2018). In the context of autonomous vehicles, it is important that the data covers different operating conditions.

As these ethical issues are ultimately left for the developers to tackle, we argue that a developer-centric approach to ethics in AI is important. While company level policies and guidelines can direct development work, micro-level decisions are nonetheless left to individual developers. Thus, developers working with AI need to be able to implement ethics into the systems they develop. This calls for both awareness of ethics among developers as well as tools and methodologies to implement it into practice. As the field of AI progresses further, the transportation industry among others becomes increasingly influenced by AVs, making these issues very current.

*2.2. Ethically Aligned Design: The Current State of Ethics in AI*

Ethics in the field of ICT has been discussed in different contexts: as a branch of professional ethics, as the application of ethical theories in ICT, and as various specific ethical issues such as Internet privacy (Bynum 2018). Ethics in the context of ICT can be understood as "the analysis of the nature and social impact of computer technology and the corresponding formulation and justification of policies for the ethical use of such technology" (Moor 1985). More recently, discussion on ethics in relation to AI systems has shifted towards what is now referred to as Ethically Aligned Design (EADe1 2019), which refers to the involvement of ethics into the design of AI and autonomous systems.

Researchers from various disciplines have recently voiced concerns over ethics in AI systems (Charisi et al. 2017). To address the growing discussion over ethics in AI, the IEEE Global Initiative on Ethics of Autonomous and Intelligent Systems was launched, which has since branded the concept of EAD. The aim of the initiative has been to raise awareness of ethical issues in AI and autonomous system design among industry professionals. Currently, the initiative has produced its first set of guidelines for EAD (EADe1 2019), and has promoted the inclusion of ethics in ICT curriculums in universities. However, for the time being, the initiative has not produced specific tools or methods that practitioners could utilize to implement EAD in their work.

The ongoing academic discussion on ethics in AI has so far converged on a set of ethical constructs for AI, some of which are also discussed in the aforementioned IEEE guidelines as key principles. The four main constructs have largely been Transparency (Dignum 2017); (Turilli & Floridi 2009), Accountability (Dignum 2017); (EADe1 2019), Responsibility (Dignum 2017), and Fairness (e.g. Flores et al. 2016), although a recent EU report on AI ethics also focused Trustworthiness (AI HLEG 2019). In our analysis, we utilize transparency, accountability, and responsibility, as well as what we argue is a subset of transparency: predictability, as a framework for our study. We illustrate one possible way of perceiving the relations of these constructs in (Fig. 1) below.

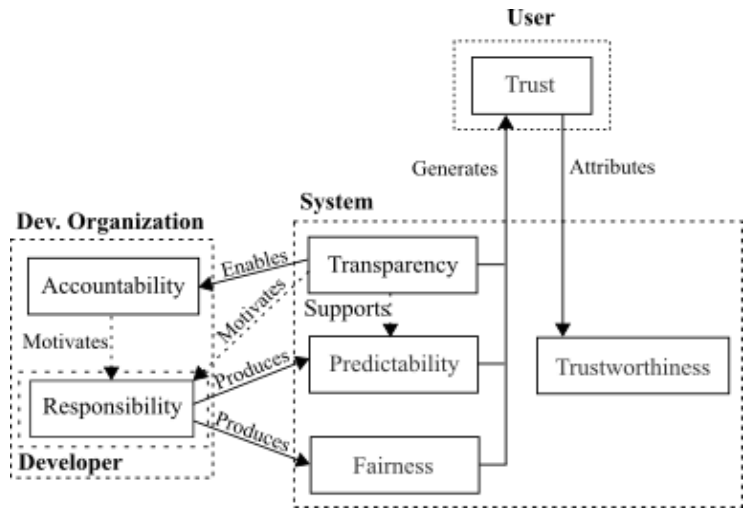

Fig. 1 Conceptualization of the Relations Between Currently Discussed AI Ethics Constructs

*Transparency* is considered currently the most important of these principles or values. Turilli and Floridi (2009) argue that transparency is the pro-ethical circumstance that makes it possible to implement ethical principles into the design process in the first place, and the EAD standard lists it as one of its key ethical principles. We consider there to be two types of transparency: (1) transparency of algorithms and data (Dignum 2017), i.e. the transparency of systems, and (2) transparency of systems development. The latter refers to e.g. what decisions were made during the development and design, why, and by whom.

*Predictability*, as the word implies, refers to whether the system acts predictably, i.e. acts how we think it should do in a given situation. It can also be seen as consistency: if an autonomous coffee machine successfully brews coffee 8 times out of 10, we are left wondering what happened the other two times and why. Predictability is, in the IEEE EAD standard, briefly discussed as a subset of transparency (EADe1 2019). While it a separate construct, transparency supports it directly: when a system is transparent, it is easier to prevent or address any issues related to predictability. We thus treat it as a subset of transparency in our data analysis.

*Accountability* and *responsibility*, while in some ways related, are also separate constructs. *Accountability* focuses on who is accountable or liable for the decisions made by the AI. Dignum (2017) in their recent works defines accountability to be the explanation and justification of one's decisions and one's actions to the relevant stakeholders. Transparency is required for accountability, as we must understand why the system acts in a certain fashion, as well as who made what decisions during development in order to establish accountability. Whereas accountability can be considered to be externally motivated, *responsibility* is internally motivated. Responsibility can be considered to be an attitude or a moral obligation for acting responsibly (EADe1 2019). In order to act responsibly, one has to weigh their options and consciously evaluate the effects of their actions and decisions.

We focus on these three main and one sub construct as we consider trustworthiness to be a higher level value that is produced by these four constructs, possibly among others. This is also the way trustworthiness is discussed in the AI report on trustworthy AI: trustworthiness is the ultimate goal of these systems (AI HLEG 2019). We argue that trustworthiness, and more specifically trust towards the system, is a feeling in the user. It is thus not possible to directly implement trustworthiness into the system as it is ultimately a subjective feeling experienced by an actor. On the other hand, e.g. transparency, and especially predictability, can be argued to produce trustworthiness.

Finally, fairness in AI ethics relates to treating all users of the systems (or those whose data is handled by the systems) equally. Fairness has, for example, been discussed from the point of view of racial bias in data handling (Greene et al. 2019); (Flores et al. 2016). While it is a relevant construct that has recently been discussed in relation to AI ethics, it is not included in the IEEE standard for EAD, and we, too, consider it to be out of the scope of this study. Discussion in the field continues, however, and we do not make an argument for or against its relevance to the ongoing discussion.

While ethics in AI has recently seen a growing focus among the academia, the current state of practice remains largely unknown. Notably, a recent study argued that the ACM Code of Ethics had had no impact on the practitioners' work practices (McNamara et al. 2018). The focus of this paper is to further explore the situation in the industry and to begin tackling the present lack of tooling for EAD, as we discuss in the next section after presenting the second part of our theoretical framework in the final subsection of this section.

2.3. *Commitment*

As the theoretical framework for this study, we approach ethics in AI through the lens of commitment. Aside from behavioral studies from the field of e.g. psychology, commitment has been studied in the past in relation to software process improvement (SPI) (Abrahamsson 2002). As we approach ethics in AI from the point of view of developers and, furthermore, take on a method, practice, and tool-focused approach, we build on these past studies on commitment in context of SPI.

Commitment is a long-standing area of research in industrial psychology and organizational behavior (Benkhoff 1997). The idea of commitment has been of interest primarily because of the assumption that the commitment of employees relates to performance. O'Reilly III and Chatman (1986) remark that "although the term commitment is broadly used to refer to antecedents and consequences, as well as the process of becoming attached and the state of attachment itself, it is the psychological attachment that seems to be the construct of common interest." Drawing from this indirect definition, we define commitment to refer to the attachment an individual feels towards an object (organization, ideal, etc.). In this study, we are interested in the commitment of software developers, and specifically AI developers, towards ethics in AI design.

In relation to commitment in the specific context of SPI, Abrahamsson (2002) proposed a model of commitment nets (Fig. 2). The model suggests that drivers, both internal and external, may result in commitment which would then manifest as actions, and those actions would then lead to both intended and potentially unintended outcomes. Using this commitment new model as the theoretical framework of this study, we focus specifically on the concerns and actions of the developers, as we will discuss in detail in the following research design section.

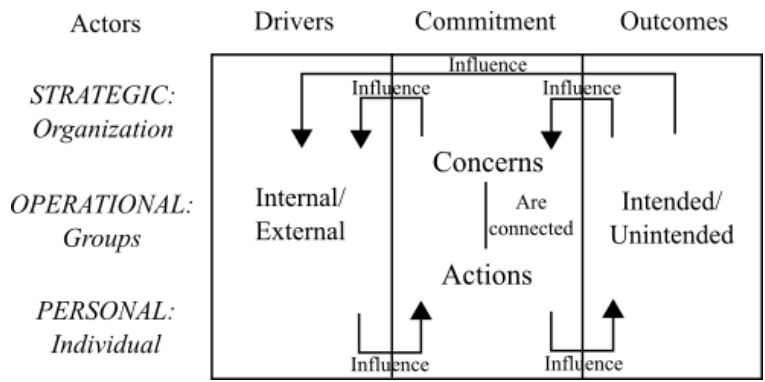

Fig. *2 Commitment Net Model (Abrahamsson 2002)*

## 3. Research Design

This study was carried out as a case study featuring five cases. Each case company develops AI systems, although in different fields, or only as a portion of their business operations, as can be seen in (Table 1) below. In each case company, we conducted semi-structured qualitative interviews focusing on ethics in AI design. The interview questions in their entirety can be found on (Vakkuri 2018).

In short, the interview protocol was designed to focus on the key constructs discussed in section 2.2: transparency, accountability, responsibility, and predictability. We avoided directly discussing ethics as different individuals have different conceptions of what ethics is in this context, as is underlined by the on-going academic discussion on the topic as well. Instead, we focused on asking practical questions related to the ethical principles such as transparency (e.g. "How well the development process is being documented? For instance, can certain functions or decisions made during the development process be led back to the individuals behind them?")

Table 1. Case Company Information

| Case | Company Description | Respondent [Reference] |
| --- | --- | --- |
| Company A | Large, >400 employees; Software, Generic | Data Scientist/Engineer [R1] Senior Data Scientist [R2] |
| Company B | SME (Small/Micro), <25 employees; Software, Healthcare | Development Lead [R3] |
| Company C | SME (Small/Micro), <25 employees; Software, Process Industry | CTO [R4] |
| Company D | Large (Multinational), >100 000 employees; Consulting | Functional Designer [R5] |
| Company E | Large (Multinational), >25000 employees; Vehicle | (AI) Development Lead [R6] |

We utilized the commitment net model of Abrahamsson (2002), which was discussed in detail in the previous section, as the theoretical framework for the analysis of these cases. We approached commitment through the *concerns* that the employees might have had towards implementing ethics in AI design, as well as through the *actions* they might have taken as a result of their concerns – if any. In doing so, we sought to understand whether any commitment towards ethics in AI design existed in the case companies. To give a practical example, if one indicates concern towards losing weight but exhibits no actions such as making dietary changes or exercising, there is no commitment present. However, rather than studying whether the case companies were committed to ethics or not, we focused on their actions by identifying practices, tools, or methods through which they had possibly addressed ethical concerns, i.e. *how* they had implemented ethics.

In our analysis of the data, we summarize our findings through what we refer to as Primary Empirical Conclusions (PEC). We consider these to be implications that are worth noting despite occasionally being outside the direct scope of our research question. They are further discussed in the discussion section.

## 4. Empirical Results

Our interviews with the case companies indicated that the industry is of the potential importance of ethics in AI. When asked "do you consider taking into account ethics useful for your organization, and if yes, how?", every respondent agreed that ethics is useful. However, despite ethics being perceived as important, the case companies had highly differing views on how relevant it was for them in practice. Indeed, in response to the question "do your AI development practices take into account ethics, and if yes, how?", all respondents answered 'no'. This underlined that the companies did not have clear tools or methods for implementing ethics.

This disconnect seemed, in part, to also stem from a lack of consensus on what ethics actually referred to in this context in the first place. This underlines a gap between the on-going academic discussion and the industry:

> "...I actually try to use the word 'ethics' as little as possible because it's the kind of word that everyone understands in their own way, and so they can feel that it's not relevant to what we're doing at all..." [R4]

> "...the discussion on AI ethics doesn't really affect most ... excluding maybe Google and some others like that ... the AI really isn't at the level where it would really necessitate in-depth ethical consideration" [R3]

> **PEC1:** Ethics is considered important in principle, but as a construct it is considered detached from the current issues of the field by developers. I.e. the on-going academic discussion on AI ethics has not permeated the industry at large.

Only the respondent involved in developing a medical AI system had a more practical view of ethics in relation to their current system. However, the respondent noted that the ethical consideration had already been carried out externally. Indeed, fields such as the field of medicine inherently have very strict regulations regarding e.g. data management, leaving little leeway for developers to make their own ethical decisions:

> "We have in-house quality measurements and these regulation requirements are very strict, so therefore these things pretty much come as a given for us. And of course if you think about it the other way, we consequently think about these things [ethics] even less because we already have such clear regulations and requirements for what we do" [R3]
>
> **PEC2**: Regulations force developers to take into account ethical issues while also raising their awareness of them

On the other hand, though ethics as a construct was considered impractical and too theoretical, the respondents did all nonetheless concern themselves with various constructs related to AI ethics (in this case: transparency, predictability, accountability, and responsibility). These constructs were considered practical by the respondents, as we will discuss next in our analysis using the commitment net model.

*4.1. Transparency*

The idea of transparency, as it is currently discussed among the academia in relation to the ethics of AI, was among the concerns of all the case companies. The companies were concerned with transparency both in the sense of transparency of systems, as well as the transparency of development. Furthermore, transparency of systems was considered both from the point of view of the developers, as well as the end users of the system. However, while these concerns were largely universal among the five case companies, the actions taken to address them varied highly between companies.

All five case companies indicated concerns related to system transparency from the point of view of developers. However, only three companies had taken clear actions to address their concerns, indicating a lack of commitment on this front (illustrated in Table 1 below):

> "The most important thing is that we can see directly how it works, and that it is, like, trackable -- at this stage, and in the future" [R5]
>
> "...it is typically a little un-transparent how the decisions [of the ML AI] are made. Of course we can analyze them, but due to the complexity of the neural network architecture, it's a little difficult to directly explain why it decided to do something." [R6]

Table 2. Commitment Towards Transparency to Developers

| Driver | Actor | Concern | Action |
|---|---|---|---|
| Project need | R1 |  | No recognized actions |
| Legislation; Regulations | R3 |  | Devoting time to understanding the training data |
| Company need | R4 | Keeping the system understandable to the developers (i.e. transparency to developers) | Devoting time to understanding the AI used as a template for the system; Building analytics into the system |
| Company need | R5 |  | No recognized actions; (Planned future action: documentation) |
| Company need | R6 |  | Devoting time to understanding the training data and testing data; Mode verification |

Whereas transparency from the point of view of developers was considered in relation to e.g. the algorithms and the neural network architecture, transparency from the point of view of the users was considered on a less technical level (Table 3). The respondents felt that the users had little reason to be able to see inside the system or the so-called black box as such. It was considered more important that the users would be able to understand how it works on the very basic level:

> "Our systems are aimed at these... operational personnel, like the paper plant guys down on the factory floor [...] they don't really know what happens inside the system and we don't feel that they really need to know, either [...] they just understand that, okay now all this data goes in, and the suggestions are then based on that data" [R4]

> "...the users won't really notice a difference compared to the earlier systems they have used. We just want to offer them better and more timely data. So that's of course one question: how to make it clear for them that there are some uncertainties there so that they don't expect the information to always be perfect. But... I don't really know how much of a problem this is -- I haven't really spoken to our end-users" [R5]

> **PEC3:** Developers have a perception that the end-users are not tech-savvy enough to gain anything out of technical system details

Table 3. Commitment Towards Transparency to Users

| Driver | Actor | Concern | Action |
| --- | --- | --- | --- |
| Project need | R1 |  | No recognized actions |
| No clear driver | R2 |  | Educating the customer/user |
| Market edge; Process improvement | R3 |  | No recognized actions |
| Company need; Professionalism | R4 | Keeping the system understandable to the end users (i.e. transparency to users) | Educating the customer/user |
| Company need | R5 |  | Writing helpful system descriptions |
| Company need; Professionalism | R6 |  | Educating the customer/user; Communication with customer/user |

Moving from transparency of systems to transparency of systems development, four of the five companies indicated clear concern towards it and had taken actions to address the concern (Table 4). Largely, (code) documentation was considered to be the primary way of producing transparency in the development process by making it apparent who made what changes, why, and when. Additionally, conducting audits was discussed as one tangible practice for producing transparency in the development process. This was one of the few areas where a consensus among the companies could be observed in ethical practices:

> **PEC4:** Documentation and audits are established Software Engineering project practices that form the basis in producing transparency in AI/AS projects

Table 4. Commitment Towards Transparency of Development

| Driver(s) | Actor | Concern | Action(s) |
|---|---|---|---|
| Project need; Customer need | R1 | | Documentation |
| Project need; Customer need | R2 | | Documentation; Conducting audits; Personnel control |
| Customer need; Market need; Regulations | R3 | Keeping track of who does and decides what and why (I.e. transparency of development) | Documentation; Conducting audits, audit trail |
| Company need | R5 | | Documentation |
| Company need | R6 | | Launch of new management process |

*4.2. Predictability*

One of the main concerns shared by all respondents was the potential unpredictability of the system (Table 5). The respondents discussed clear actions they had taken to either avoid unpredictable behavior, to mitigate it, or to prevent it in the future in case it takes place. An example of such an action can be ML management by means of using different sets of training data or limiting its utilization.

> "...we have even cut some functionalities [...] of the system in order to make it more predictable, which has reduced the amount of unexplained results we have gotten out of it [...] in practice we've been able to explain all of the faulty results so far" [R3]

**PEC5:** Machine learning is considered to inevitably result in some degree of unpredictability. Developers need to explicitly acknowledge and accept heightened odds of unpredictability.

Table 5. Commitment Towards Preventing Unpredictability

| Driver(s) | Actor | Concern | Action(s) |
|---|---|---|---|
| No clear driver | R1 | System acts unpredictably (i.e. preventing an incident) | Awareness of unpredictability; Recognizing what errors are acceptable; Preparedness for incidents of unpredictability |
| Company need | R2 | | Representative training data; Training for designer |
| No clear driver | R3 | | Reduce functionalities and complexity of system; Narrow the scope of use of machine learning |
| No clear driver | R4 | | Accept the (minimal) odds of unpredictability; Acknowledging that statistical tools also make mistakes; Root cause analysis |
| No clear driver | R5 | | Using the system only in confined spaces |
| Company need | R6 | | AI/ML model validation |

In discussing steps they had taken to avoid unpredictability, the respondents also discussed their concerns related to a hypothetical situation in which the system has already acted unpredictably (Table 6). All six respondents and five case companies had outlined some courses of action for such a scenario, although some of the actions pointed towards a lack of commitment (e.g. apologize and react on a case-by-case basis is a very ad hoc plan).

Table 6. Commitment Towards Addressing an Incident of Unpredictability

| Driver | Actor | Concern | Action |
|---|---|---|---|
| Customer need; Company need | R1, | System makes mistake in production (i.e. hypothetical scenario in which an incident took place) | Accept the (minimal) odds of unpredictability; Be willing to react; Apologize |
| Company need; Project need; Professionalism | R2 | | Be willing to react; Apologize; [Planned future action: communication/action plan] |
| Customer need; Financial | R3 | | Feedback options to product development; Using mistake as example in learning data; Accept the (minimal) odds of unpredictability; Acknowledging that statistical tools also make mistakes |
| No clear driver | R4 | | Piloting before full release; Reacting feedback and fixing issues; Narrowing functionalities in design |
| Company need; Customer need | R5 | | Piloting oversite; Cutting system functionalities; Fixing bugs when noticed |
| Company need; Customer need; Legislation | R6 | | Backup systems |

Finally, in relation to predictability, four of the respondents discussed cyber security threats as a part of unpredictable system occurrences (Table 7), even if they are caused by external actors as opposed to the system itself. Indeed, however, in the case of especially CPSs, cybersecurity threats can pose life-threatening danger if e.g. an autonomous bus is hijacked digitally, and thus preventing these threats is pivotal from the point of view of ethics as well. Given that cybersecurity is a longstanding area of research and industry practice, companies generally have established policies and even cybersecurity departments for dealing with cybersecurity issues. Thus, few actionable measures or practices were underlined by the respondents in response to their actions in tackling cybersecurity concerns.

Table 7. Commitment Towards Cybersecurity

| Driver(s) | Actor | Concern | Action(s) |
| --- | --- | --- | --- |
| Company need; Customer need | R1 | | Follow quality process and corporate policy |
| Company need; Project need; Professionalism | R2 | Cybersecurity / Data security / Adversary attacks | Recommendations on how to prepare; Awareness of context of use (i.e. who can do and what with the system) |
| Company need; Customer need; Legislation | R3 | | Follow quality process and corporate policy |
| Company need; Customer need | R6 | | Backup systems; Preparing for attacks |

*4.3. Accountability and Responsibility*

The consensus among the respondents was that no system could be completely fault-free, with five respondents expressing concern towards potential harm caused by their system(s) (Table 8). Most respondents could also list some actions their organization had taken to either avoid or mitigate harm caused by their system. However, some of the respondents felt that their system(s) had no direct potential for harm even if it did act unpredictably or wrongfully, due to it e.g. being a purely digital business intelligence system.

> **PEC6:** Developers consider the harm potential of a system primarily in terms of physical harm or harm towards humans. Potential systemic effects are often ignored.

Additionally, the respondent working on healthcare AI (R3) indicated a more personal approach to responsibility than the other respondents because they felt that they were directly responsible for the well-being of some of their users.

> **PEC7:** Physical harm potential motivates personal drivers for responsibility

Notably, the respondents ultimately outsourced the responsibility and/or accountability to their users despite exhibiting a commitment to mitigate or prevent harm. They felt that they had taken what measures they could to prevent harm, and that it was then up to the user to stay safe (e.g. doctors need to make informed decisions based on the data):

> **PEC8:** Main responsibility is outsourced to the user, regardless of the degree of responsibility exhibited by the developer

Table 8. Commitment Towards Responsibility for Potential Harm

| Driver(s) | Actor | Concern | Action(s) |
|---|---|---|---|
| Customer need | R1 | | Adhering to contracts; Responsible project management |
| Company need; Project need; Personal | R2 | | No recognized actions |
| Personal | R3 | Responsibility for potential harm caused by the system or a specific algorithm | Accept the (small) odds of harm; Communication with the customer to minimize the risk of harm |
| No clear driver | R5 | | Design the system in such a way that even wrong decisions are not harmful |
| No clear driver | R6 | | Minimizing potential harm; Accept the (small) odds of harm; Build a system that produces less harm than humans in the same context |

As the respondents discussed having concerned themselves and their project teams very little with direct discussions about ethical matters related to their systems, they did not consider responsibility strongly from an ethical point of view. Instead, they approached responsibility largely from the point of view of delivering a product that fulfilled expectations set for it (Table 9) internally, by various stakeholders, or by regulations. Some of the respondents also felt that delivering a quality product was their responsibility as professionals of the field.

**PEC9:** Developers typically approach responsibility pragmatically from a financial, customer relations, or legislative point of view rather than an ethical one.

Table 9. Commitment Towards Delivering a Good System

| Driver(s) | Actor | Concern | Action(s) |
|---|---|---|---|
| Company need; Commercial; Professionalism | R1 | | Setting realistic goals for the system |
| Commercial | R3 | | No recognized actions |
| Company need; Customer need; Professionalism | R4 | Delivering a working product / Delivering what was promised | Piloting; Keeping the human in the loop |
| No clear driver | R5 | | Discussion inside project team; Communication with customer |

## 5. Discussion

We have collected the Primary Empirical Conclusions (PECs) discussed in the results section into Table 10 below. For the purpose of this section, they have been split into three categories based on their contribution: (1) empirically validates existing literature, (2) contradicts existing literature, and (3) new knowledge. On a general level, the primary contribution of this study is its novel viewpoint on AI ethics: the developers and the state of practice. The existing literature in the field of AI ethics is lacking in empirical studies and has so far focused largely on discussion among the academia.

Table 10. Primary Empirical Conclusions of the Study

| PEC | Theoretical component | Description | Contribution |
|---|---|---|---|
| 1 | Conceptual | Ethics is considered important in principle, but as a construct it is considered detached from the current issues of the field by developers. | Empirically validates existing literature |
| 2 | Conceptual | Regulations force developers to take into account ethical issues while also raising their awareness of them. | Empirically validates existing literature |
| 3 | Transparency | Developers have a perception that the end-users are not tech-savvy enough to gain anything out of technical system details. | Contradicts existing literature |
| 4 | Transparency | Documentation and audits are established Software Engineering project practices that form the basis in producing transparency in AI/AS projects. | Empirically validates existing literature |
| 5 | Predictability | Machine learning is considered to inevitably result in some degree of unpredictability. Developers need to explicitly acknowledge and accept heightened odds of unpredictability. | Empirically validates existing literature |
| 6 | Responsibility & Accountability | Developers consider the harm potential of a system primarily in terms of physical harm or harm towards humans. | New Knowledge |
| 7 | Responsibility & Accountability | Potential systemic effects of AI systems, as well as their social and emotional impacts, are ignored by developers. | New Knowledge |
| 8 | Responsibility & Accountability | Physical harm potential motivates personal drivers for responsibility. | Empirically validates existing literature |
| 9 | Responsibility & Accountability | Main responsibility is outsourced to the user, regardless of the degree of responsibility exhibited by the developer. | New knowledge |
| 10 | Responsibility & Accountability | Developers typically approach responsibility pragmatically from a financial, customer relations, or legislative point of view rather than an ethical one. | New knowledge |

The key finding of this study is that there is a gap between research and practice in the field of AI ethics (PEC1). The academic discussion on AI ethics and the values related to it (transparency etc.) seems to not have affected the industry yet. This is consistent with the findings of McNamara et al. (2018) who concluded that the ACM Code of Ethics (Gotterbarn et al. 2018) had done little to change the way developers work. This gap has also been discussed in the most recent version of the IEEE EAD guidelines (EADe1 2019), although without direct empirical evidence.

We argue that this gap largely stems from a lack of tooling and methodologies in the area, as has been suggested by Whittlestone et al. (2019) as well. Based on our data, industry professionals currently address ethical issues through highly differing ad hoc methods. While guidelines such as the IEEE EAD guidelines (EAD1e) and guidelines made by industry experts exist (e.g. Google AI principles (Pichai 2018)), they are not actionable (Whittlestone et al. 2019). The guidelines discuss principles and values but offer little help in implementing them into practice. Tools and methods are needed to help practitioners make use of these guidelines.

Aside from tooling, one way of addressing this gap would be through changes in legislation and the implementation of regulations (PEC2). However, legislative changes are slow and may struggle to keep up with the advances in technology. They may also have negative, limiting effects on AI development (e.g. regulations on international waters limit testing maritime AVs (One Sea Ecosystem 2017).

Developers currently do not approach ethics in a systematic manner and do not utilize any tools or methodologies to implement it. However, ethical values discussed in academic literature are nonetheless taken into account in the industry to some extent. According to the IEEE EAD guidelines (EADe1 2019), documentation is key in producing transparency. This was also acknowledged by all case companies (PEC4), although the sufficiency of their documentation remains unknown. Similarly, the challenges ML poses to system predictability are discussed in existing literature and also acknowledged by industry professionals (PEC5).

On the other hand, while the IEEE EAD guidelines (EADe1 2019) encourages transparency in terms of providing users with technical details of the systems as well, developers feel that their users do not possess the technical knowledge to make any use of said information (PEC3). Here the opinions of the developers also notably contradict existing literature in which transparency has been extensively discussed e.g. from the point of view of the users or the general public being able to understand the technical side of the system. The developers were not necessarily averse to the idea of letting the users "look inside the system" but felt that there was little reason to let them do so.

In terms of responsibility, developers currently do not possess the skills to evaluate the harm potential of AI systems comprehensively. Developers seem to exhibit a narrow view of the harm potential of such systems, focusing on physical harm (PEC6). This is a topic that has not been extensively studied thus far but practical incidents such as the Cambridge Analytica one do point towards this being the case, and have raised awareness of this issue. I.e. either developers are unaware of these issues or they are simply ignored e.g. in favour of completing the tasks they are assigned.

When a system is considered to have physical harm potential, developers also feel more strongly about acting responsibly (PEC8). On the contrary, the social and emotional impacts of AI systems are ignored (PEC7). Developers also do not consider the systemic effects of AI systems, while in reality they can be important (German Federal Ministry of Transport and Digital Infrastructure 2017). E.g. AVs do not only affect their passengers; drivers at large may act differently in the presence of autonomous vehicles out of caution, which in turn may confuse the algorithms of the AVs. This further underlines the gap between research and practice in the area, as AI ethics literature discusses the harm potential of AI systems extensively through e.g. social issues such as racial bias (Albarghouthi and Vinitsky 2019); (Flores et al. 2016).

In this vein, however, we feel that we cannot expect developers to do such comprehensive ethical analysis unassisted and without training on the matter. Carrying out such analyses calls for distribution of work in organizations, or even hiring ethical experts to carry out the analysis. Furthermore, we once more underline the importance of tools and methods in this regard. The IEEE EAD initiative has also begun to address this issue by advocating for the inclusion of AI ethics into university curricula, mainly in the US.

On a further on responsibility, developers seldom consider responsibility important purely for ethical reasons. Rather than being concerned about being ethical, they are concerned about potential financial losses or bad publicity resulting from the system being unethical (PEC10). This is to some extent similar to how companies have approached environmental issues or business ethics at large, although nonetheless new in the specific context of AI ethics. Companies are likelier to tackle these issues for financial or legislative issues, as opposed to simply acting responsibly. This should be considered when attempting to raise awareness of AI ethics in the industry.

Regardless of the degree of responsibility exhibited by the developers, the responsibility is ultimately outsourced to the user(s) of the system (PEC9). I.e. the developers feel that the user should always be critical towards the suggestions of the system, whether the user is a doctor or a factory worker, and that how they use the system is their responsibility. Similar lines of argumentation are seen in relation to e.g. firearm legislation and thus while this is new in the context of AI ethics, outsourcing responsibility in this sense as a phenomenon is not novel.

Moreover, outsourcing responsibility in this context is interesting when combined with PEC3, as the developers simultaneously feel that their end-users are not tech savvy enough to benefit from being explained or shown the technical details of the system. Yet, despite the users thus having no in-depth understanding of how the systems work, the developers feel that the users should be able to evaluate the actions of the systems in an informed fashion. This issue has been, in part, acknowledged in existing literature. Scholars have repeatedly voiced their concerns over black boxes and demanded *explainable* AI systems. (Bryson and Winfield 2017); (Adadi and Berrada 2018).

Finally, in terms of future research directions, we suggest three. First, research addressing the ongoing discussion on AI ethics should continue until a better consensus on terminology is reached. While transparency and accountability can be argued to be universally accepted constructs in AI ethics, constructs such as fairness are still emerging. Secondly, we urge future research to focus on tackling the gap between practice and research in the area through the development of tooling and methodologies that could be used to implement AI ethics into practice. Thirdly, further studies on the current state of practice from the point of view of practices can aid the creation of said methods and tools through the discovery of existing good practices (e.g. PECs 4 and 5)

*5.1. Limitations of the Study*

Though our findings are based on a multiple case study of five companies, we nonetheless underline the limitations of qualitative data. Given the qualitative approach of the study, we cannot ascertain that our findings are indeed representative of the current state of the industry at large. A further limitation in our data is that all five case companies were either Finnish or international companies whose Finnish branch was the only one involved in the study. This is a limitation in this context specifically because much of the discussion on AI ethics has been US-based, and many of the AI ethics university courses motivated by the EAD initiative are also US-based. It is therefore possible that especially US companies might be more aware of the academic discussion on AI ethics and that the state of practice in these companies may be different.

## 6. Conclusions

In this paper, we have conducted a case study to understand the current state of practice in relation to ethics in AI. The case study featured five case companies, in which the data was gathered through semi-structured, qualitative interviews. We utilized the commitment net model to analyze the data through the concerns the organizations or individuals exhibited towards various ethical issues, as well as the actions they had taken to address said concerns.

In summary, developers consider ethics important in principle. However, they consider ethics as a construct impractical and distant from the issues they face in their work. There is thus a clear gap between research and practice in the area as the developers are not aware of the academic discourse on the ethics of AI.

The key finding of this study was that none of the case companies utilized any tools or methodologies to implement AI ethics. Based on our data, developers lack ways to systematically implement ethics into practice. They tackle ethical issues separately and in an ad hoc fashion, using highly differing practices across organizations. While various guidelines for AI ethics currently exist, written by both practitioners and scholars alike, these guidelines are not used by industry experts. Indeed, they consist of principles and values rather than actionable practices.

We thus recommend that future research in the area seek to: (1) develop tools and methodologies to help industry experts implement AI ethics into practice, and (2) help reach a consensus in the ongoing conceptual discussion on AI ethics so that the tools and methods can utilize a stable and agreed-upon set of constructs.